\documentclass[aps,pra,twocolumn,showpacs,superscriptaddress,letterpaper,longbibliography]{revtex4-1}

\usepackage{graphicx}
\usepackage{mdframed}
\usepackage{framed}
\usepackage{indentfirst}
\usepackage{natbib}
\usepackage{amsmath,amssymb}
\usepackage{subfigure}
\usepackage{enumitem}
\usepackage{epstopdf}

\begin{document}

\title{An alternative model for administration and analysis of research-based assessments}


\author{Bethany R. Wilcox}
\affiliation{Department of Physics, University of Colorado, 390 UCB, Boulder, CO 80309}

\author{Benjamin M. Zwickl}
\affiliation{School of Physics and Astronomy, Rochester Institute of Technology, Rochester, NY 14623}

\author{Robert D. Hobbs}
\affiliation{Department of Physics, Bellevue College, 3000 Landerholm Cr SE, Bellevue, WA, 98007}

\author{John M. Aiken}
\affiliation{School of Physics, Georgia Institute of Technology, 830 State Street, Atlanta, GA 30332}

\author{Nathan M. Welch}
\affiliation{Department of Physics, University of Colorado, 390 UCB, Boulder, CO 80309}

\author{H. J. Lewandowski}
\affiliation{Department of Physics, University of Colorado, 390 UCB, Boulder, CO 80309}
\affiliation{JILA, National Institute of Standards and Technology and University of Colorado, Boulder, CO 80309}

\begin{abstract}
Research-based assessments represent a valuable tool for both instructors and researchers interested in improving undergraduate physics education.  However, the historical model for disseminating and propagating conceptual and attitudinal assessments developed by the physics education research (PER) community has not resulted in widespread adoption of these assessments within the broader community of physics instructors.  Within this historical model, assessment developers create high quality, validated assessments, make them available for a wide range of instructors to use, and provide minimal (if any) support to assist with administration or analysis of the results.  Here, we present and discuss an alternative model for assessment dissemination, which is characterized by centralized data collection and analysis.  This model provides a greater degree of support for both researchers and instructors in order to more explicitly support adoption of research-based assessments.  Specifically, we describe our experiences developing a centralized, automated system for an attitudinal assessment we previously created to examine students' epistemologies and expectations about experimental physics.  This system provides a proof-of-concept that we use to discuss the advantages associated with centralized administration and data collection for research-based assessments in PER.  We also discuss the challenges that we encountered while developing, maintaining, and automating this system.  Ultimately, we argue that centralized administration and data collection for standardized assessments is a viable and potentially advantageous alternative to the default model characterized by decentralized administration and analysis.  Moreover, with the help of online administration and automation, this model can support the long-term sustainability of centralized assessment systems.  
\end{abstract}

\maketitle

\section{\label{sec:intro}Introduction}

One of the significant contributions of the physics education research (PER) community has been the creation of a number of research-based, standardized assessments (see Ref.\ \cite{PhysPortwebsite} for an extensive list).  Research-based assessments (RBAs) include (but are not limited to) both concept inventories and instruments targeting students' epistemologies.  These assessments are typically based on known student ideas, validated through expert and student review, pilot tested to demonstrate statistical validity and reliability, and administered for the purpose of low-stakes formative assessment \cite{adams2011development,wilcox2015assessments}.  RBAs provide a standard and valid measure of student outcomes that can be compared across semesters, instructors, institutions, and pedagogies.  Such measures help to quantify the relative success of educational strategies, and thus can have significant impact on the initial adoption and continued use of new innovations by individual instructors, as well as at the programmatic level \cite{henderson2014assessment, seymour2002change, chasteen2015sei}.  Additional discussion of the importance and value of assessment, including the use of RBAs, can be found in Refs.\ \cite{henderson2014assessment, madsen2016rba}.  

Despite the utility and potential impact of RBAs, previous work suggests that these assessments are not being utilized by undergraduate physics instructors as commonly as members of the PER assessment community might expect or want.  For example, Henderson \emph{et al.} \cite{henderson2014assessment} interviewed 72 physics faculty on their use of various assessment strategies and found that only 33\% reported using RBAs in their courses.  To explore the question of why RBAs are not being adopted by a larger portion of the broader community of physics instructors, we draw from literature on promoting and sustaining adoption of educational innovations.  This work argues that, in addition to designing innovations that meet the needs and goals of potential users, successful adoption of innovations like RBAs also depends on two distinct processes: ``dissemination'' (i.e., spreading the word), and ``propagation'' (i.e., promoting sustained adoption) \cite{stanford2016adoption}.  Dissemination focuses on what the developer does whereas propagation focuses on the users of educational innovations \cite{khatri2016adoption}.  

\begin{figure*}
\includegraphics[width=1\linewidth]{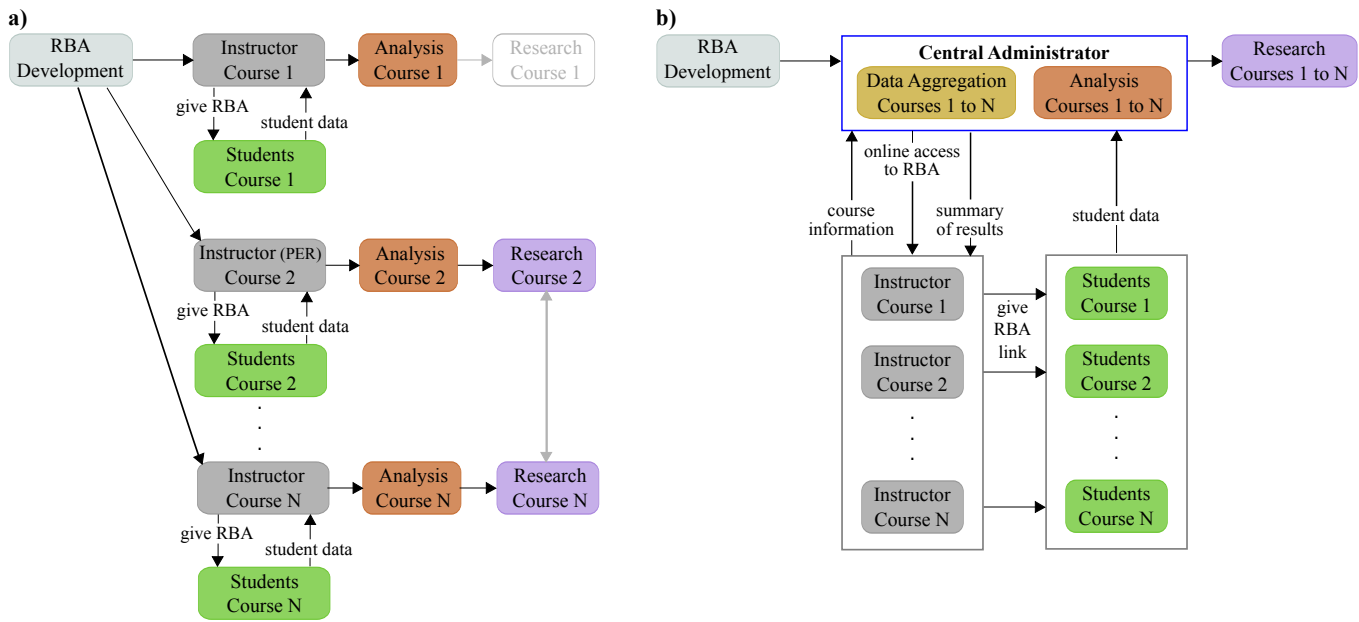}
\caption{a) A schematic representation of the default model for the propagation of RBAs that has been used historically in PER. Here, light grey arrows denote rare or non-existent connections.  This model depends almost entirely on the instructor to drive the administration and analysis of their students' results.  There is also little to no mechanism for data from multiple courses to be centrally aggregated or compared. b) A schematic representation of an alternative, centralized model for the administration and analysis of RBAs. This model depends almost entirely on a centralized administration system to facilitate administration and analyze students' responses.  This model also creates a centralized pool of comparison data that can be used for research and analysis purposes. }\label{fig:models}
\end{figure*}

Historically, RBA developers have focused almost exclusively on dissemination through strategies such as presentations at professional conferences, publications in peer-reviewed journals, and online access to the instrument for interested instructors \cite{henderson2011change, singer2012dber}. These historical dissemination strategies focus on informing physics instructors that RBAs are useful and available.  However, beyond the creation of comprehensive ``How to'' guides and scoring resources (e.g., electronic scoring spreadsheets) \cite{PhysPortwebsite, EM1website}, RBA developers have rarely, if ever, explicitly addressed issues of propagation.  This failure to explicitly attend to propagation is consistent with an implicit assumption that the only necessary ingredient for the sustained adoption of RBAs is to make physics instructors aware of their existence and value \cite{henderson2011change, stanford2016adoption, seymour2002change}.  The result of this assumption is a default model for the propagation of RBAs.  In this model, individual instructors are solely responsible for administering the RBA, as well as analyzing and interpreting their students' performance.  This process is typically done with minimal, if any, direct support from the assessment developer.  A schematic representation of this default propagation model is given in Fig.\ \ref{fig:models}a.

The default model for RBA administration and analysis fails to account for important factors that can potentially impact successful propagation.  For example, the dynamics of the broader instructional system, which includes the instructor, the department, and the institution, can encourage or discourage instructors from utilizing RBAs \cite{stanford2016adoption}.  To better understand these additional factors that contribute to instructors' use (or not) of RBAs, Madsen \emph{et al.} \cite{madsen2016rba} interviewed physics faculty and department heads who were familiar with RBAs about their experiences with, and perceptions of, these instruments.  They identified a variety of issues around faculty use of RBAs, not all of which can be overcome through improved propagation strategies (e.g., misalignment of the RBA with faculty learning goals).  However, lack of support during the process of utilizing RBAs and interpreting the results was cited as a barrier by many instructors \cite{madsen2016rba}.  In particular, instructors encounter challenges with the logistics of administering RBAs and want help with this process (e.g., information on best practices for giving tests, tools to automate the collection and analysis of RBA data). 

Challenges that instructors encounter when administering and analyzing RBAs are not meaningfully acknowledged in the default model RBA propagation (Fig.\ \ref{fig:models}a).  Instead, it is assumed that individual instructors will successfully administer, score, and analyze these assessments independently, despite the fact that many instructors report encountering difficulties in this process.  Moreover, Madsen \emph{et al.} \cite{madsen2016rba} found that instructors often face additional hurdles as they attempt to productively interpret RBA data from their course (e.g., interpreting results in courses with few students, identifying actionable implications for teaching).  In particular, instructors are often interested in how their students' responses compare to those of students in similar courses nationally \cite{madsen2016rba}.  However, collection and analysis of RBA data has historically been localized to individual faculty, thus appropriate comparison data is often difficult to find.  This lack of any clear mechanism for accessing and/or aggregating comparison data severely limits one of the main affordances of RBAs -- providing a standardized measure of student learning across classroom formats, curricular variations, pedagogical strategies, and institutions.  

Additionally, the localization of the collection and analysis of RBA data to individual faculty in the default model for the propagation of RBAs also has consequences for PER community.  Unless the instructor is, or is collaborating with, a physics education researcher with Institutional Review Board  approval to conduct human subjects research, students' responses are rarely made available for research purposes.  In cases where data from RBAs are used for research purposes, the resulting research is often highly contextualized to a single course or institution.  Additionally, research contexts are often not captured and reported in a systematic or consistent way across multiple studies.  This contextualization, along with the lack of standardized administration, analysis, and reporting of results from RBAs, can make it more difficult for researchers to identify and provide valid comparison data from similar courses.  Significant contextualization also places significant limitations on the types of questions that can be tackled by researchers.  Without centralized data from multiple courses and institutions along with standardized information about course context, it often becomes logistically impractical to answer questions about, for example, transferability of effective course transformations, or generalizability of student difficulties.  

In recent years, several groups in the PER community have made explicit attempts to address the challenges inherent in the default model for the propagation of RBAs through an alternative model that is characterized by centralized administration, data collection, and analysis.  Examples of this kind of centralized administration of standardized assessments can also be found in several other disciplines \cite{seymour2000salg,weston2015urssa, bonham2000wba} and contexts \footnote{Systems involving centralized data collection with comparisons to national data sets are also common in, for example, online homework systems.  These systems are not discussed here as they are of a different nature and serve a different purpose than the RBAs used within PER.  They are proprietary systems used for high-stakes, individual assessment and are rarely used for research purposes. }.  The goal of the remainder of this paper will be to provide a proof-of-concept for this centralized RBA model within PER by presenting, in detail, the design and implementation of a newly launched, automated online system created to facilitate the administration and analysis of the Colorado Learning Attitudes about Science Survey for Experimental Physics \cite{zwickl2014eclass} (E-CLASS, Sec.\ \ref{sec:eclass}).  Additionally, we also briefly discuss related efforts by several other research teams (Sec.\ \ref{sec:other}), and end with discussion of the characteristics, advantages, and challenges associated with a centralized administration and data collection model for the propagation of RBAs in PER (Sec.\ \ref{sec:discussion}).

\section{\label{sec:model}An alternative model}

To address many of the shortcomings of the default model for the propagation of RBAs, we propose an alternative model that localizes administration, data collection, and analysis to a single centralized system (Fig.\ \ref{fig:models}b).  In this model, instructors interested in a particular RBA provide information about their course to a central administrator.  The administrator then provides online access to the RBA for the instructor and their students.  Students' responses to the RBA from all courses are then collected and analyzed directly by this central administrator.  The administrator uses the aggregate data from all courses to produce individualized reports for each instructor summarizing their students' responses relative to similar courses in the data set.  

This centralized model for the propagation of RBAs (Fig.\ \ref{fig:models}b) addresses many of the factors identified by Madsen \emph{et al.} \cite{madsen2016rba} as barriers to instructors interested in using RBAs.  The centralized administrator not only helps to reduce the burden on individual instructors by dealing with many of the logistical aspects of administering RBA, but also provides a concrete point of contact who can respond to instructor questions and provide targeted resources.  Additionally, by aggregating data from multiple courses and tagging them with course-specific meta-data (e.g., class size, level, type of instruction), the central administrator can provide appropriate comparison data that can help an instructor determine how their course compares to similar courses.  This aggregate data can also be used by the administrator for research purposes, as well as to feed back into improving the analysis that is ultimately reported to instructors.

\subsection{\label{sec:eclass}An example from the E-CLASS}

To more clearly illustrate this centralized RBA model, this section presents a concrete example of this model in practice.  This example centers around a newly automated administration system for the Colorado Learning Attitudes about Science Survey for Experimental Physics (E-CLASS) to multiple laboratory courses.  Note that while this section includes some background information of the E-CLASS itself, the focus here is on the E-CLASS administration system.  Discussion of the E-CLASS as a valid assessment instrument has been reported previously (see Refs \cite{zwickl2014eclass, wilcox2016eclass}).  The goal of this section is to provide sufficient detail on the history, development, and maintenance of the E-CLASS system both to demonstrate the model and to provide a base from which interested developers of RBAs could replicate or build off this model in the future.  

\subsubsection{\label{sec:background}The E-CLASS assessment}

The E-CLASS was developed by researchers at the University of Colorado Boulder (CU) \cite{zwickl2014eclass} to support ongoing, local and national initiatives to improve laboratory instruction \cite{zwickl2013adlab}.  The E-CLASS is a 30 item, Likert-style survey designed to measure students' epistemologies and expectations regarding the nature of experimental physics.  Items on the E-CLASS feature a paired question structure in which students are presented with a statement (e.g., ``Calculating uncertainties helps me understand my results better.'') and are asked to rate their level of agreement on a 5-point Likert scale both from their personal perspective and that of a hypothetical experimental physicist.

The E-CLASS is generally administered online and typically outside of class time.  Instructors have historically been recruited to use the E-CLASS with their students through a variety of means including, personal communication, emails to professional email lists, presentations, and publications \cite{zwickl2014eclass, zwickl2012eclass, wilcox2015convergentECLASS}.  Information about the instrument is also available online \cite{ECLASSwebsite}.  The instrument was initially developed and validated at CU, but has now been validated using data from more than 70 courses at roughly 45 institutions \cite{wilcox2016eclass}.  

\subsubsection{\label{sec:function}The E-CLASS system}

To administer the E-CLASS via our automated system, an interested instructor first completes our Course Information Survey (CIS).  Questions on the CIS collect logistical information (e.g., course start date, number of student enrolled, department demographics), as well as information about course learning goals, equipment, and pedagogy.  Logistical information from the CIS feeds into the automated system described in Sec.\ \ref{sec:cost} where responses to particular questions are used to generate unique pre- and post-survey links for each course.  The system then emails these links to the instructor at the beginning and end of the course respectively.  Each link is activated and deactivated automatically by the system based on dates selected by the instructor on the CIS.  In order to utilize student responses to the E-CLASS for research purposes, we have approval from the CU Institutional Review Board to conduct human subjects research using this online system.  Consistent with the requirements of that approval, the survey link also includes a consent form notifying students that data from the assessment may be used for research purposes and providing instructions for how to withdraw participation.  

After the pre- and post-instruction surveys have closed, the system automatically emails the instructor with a list of the names and ID numbers of all students who completed each survey, which the instructor can use to provide a small amount of participation credit.  Due to constraints around performing human subjects research, instructors never receive identifiable, raw student responses to the survey.  Instead, student responses are fed into report generation software, which creates an individual, online report for each course.   These aggregate reports summarize students' responses to the E-CLASS both overall and by-item, and also include comparison statistics pulled from our growing data set composed of student responses from previous semesters.  While generally intended to be shared between instructors, course reports are only accessible via a unique and idiosyncratic link delivered directly to the individual instructor, so an individual instructor can choose to keep the report for their class private.  

As of the fall semester 2015, the E-CLASS system has been almost fully automated, requiring only minimal input from a human administrator from the initial completion of the CIS through to the distribution of the final reports.  However, the E-CLASS has been centrally administered since its initial development in fall of 2012.  Prior to the development of the automated system, survey generation, management, and the report generation was done by-hand by a part-time administrator.  While managing the system by hand was successful in the short term, the personnel requirements of this method likely make it unsustainable over long time scales.

\subsubsection{\label{sec:cost}Developing and maintaining the E-CLASS system}

The most demanding aspect of the E-CLASS system, both financially and in terms of expertise, was developing and implementing the automation.  As indicated previously, automation is not strictly necessary in order to run a centralized assessment system, at least in the short term.  The E-CLASS was centrally administered for more than 5 semesters before the system was automated.  However, operating the system without automation required a dedicated administrative assistant who was responsible for generating, managing, and analyzing pre- and post-surveys for 40-60 courses each semester.  The automated system has reduced the administrative requirements to checking the system periodically to ensure it is functioning and responding to email requests from participating instructors.  This reduction in the need for dedicated personnel greatly increases the sustainability of this system in future semesters when grant funding for the development of the E-CLASS has ended.  

The program that governs the automation of the E-CLASS system was written in the programming language Python \cite{p2015website}.  Python was selected over a more formal relational database \footnote{A relational database is a description of tables of data and how they relate to each other} for several reasons.  Python is well known in physics, and many physicists have a basic knowledge of Python or a similar coding language.  This fact increases the sustainability of the program as future personnel will need to be familiar with both physics content and basic Python programming.  Python also includes a number of powerful statistical and representational packages that can facilitate data analysis \cite{scipy2001,numpy2011,pandas2010,matplotlib2007}.  One of the authors, an undergraduate computer science major familiar with the Python programming language, was hired to help design and then write the code for the survey automation.  The E-CLASS surveys are hosted on a commercial survey platform known as Qualtrics \cite{q2015website}.  Qualtrics is CU's official survey platform, and the university provides access to this service for all faculty and staff.  The E-CLASS automation was designed to interface with the Qualtrics application programming interface (API) in order to allow the system to automatically generate, activate, deactivate, and pull results from surveys on Qualtrics.  Additionally, hosting the surveys on Qualtrics maintains the security of the E-CLASS items as Qualtrics surveys are only accessible via their unique survey links and cannot be accessed by Googling ``E-CLASS'' or the question prompts.

In addition to the undergraduate student who produced the code for the automation, one of the authors was brought on as an outside contractor with both PER and programming experience to create the report generation software, also written in Python.  The report generator takes students' raw responses from Qualtrics and exports a static HTML report for each course, which can be viewed and shared via the web or printed.  Reports include aggregate statistics on the overall E-CLASS score, average item scores, and student demographic data for both the course and similar-level courses from previous semesters, as well as a brief description of how the data were analyzed.  The decision to generate static, rather than dynamic, reports was made primarily to enhance their sustainability.  A web application capable of generating dynamic reports with continuously updating comparison data or interactive graphs would require real-time developer support should the report system ever encounter a problem.  Alternatively, static reports, once generated, can be made available indefinitely and would not be affected should the report generator encounter an unforeseen problem.  The automation, report generator, and raw student responses are all stored and run on a local, dedicated and secure server purchased for this purpose.  Local university IT personnel assisted with the process of purchasing, setting up, and maintaining the E-CLASS server.  

As the major deliverable from the E-CLASS system, the final reports have gone through multiple iterations of review and refinement to ensure that they are clear and useful to instructors.  An example E-CLASS report can be accessed from Ref.\ \cite{ECLASSwebsite}.  To investigate instructors' perceptions of the reports, we solicited feedback from instructors who had received E-CLASS reports in previous semesters.  Nine of these instructors also participated in phone interviews in which they were asked to discuss any positive, negative, and/or confusing aspects of the report.  Interviews were conducted by one of the authors and typically lasted between 15-30 minutes.  All nine interviewees made positive comments about the E-CLASS and/or the final report.  For example, a representative quote from one of the interviewees was, ``The feedback that I get about the statistics about my students is very well laid out.  There is a very good explanation at the beginning on how to look at the data and interpret the graphs, and it is well organized and systematic and easy to follow.''  The presentation of individual course scores side-by-side with significant comparison data from other courses nationally was also mentioned as one of the major useful features of the E-CLASS report.  Additionally, interviewees noted that the reports contained a large amount of information and that the visual representations of students scores and shifts helped to facilitate interpretation.  

Despite the instructors' overall positive feedback about the report, they made a number of suggestions for improvement or clarification.  Many of these suggestions have since been incorporated into the report generation software.   For example, the ordering convention for data points in several of the graphs was modified to support greater consistency between individual reports.  This was in response to comments from instructors that suggested the original ordering convention made it difficult to compare graphs across distinct courses and/or semesters.  Several instructors also requested that the report include documentation of the accepted ``expert-like'' response for each question.  This information has now been incorporated into a separate ``Questions'' tab in the online reports.

\subsection{\label{sec:outcomes}Initial impacts of the E-CLASS system}

In this section, we discuss ways in which the centralized administration system for the E-CLASS has supported research efforts and faculty use of the E-CLASS.  

\subsubsection{Research with growing national dataset}

Over the previous six semesters during which the E-CLASS has been centrally administered, we have collected over 5200 matched pre- and post-instruction student responses from 120 distinct courses spanning 64 institutions across the United States.  These courses include introductory labs for non-majors up through advanced laboratory courses for senior physics majors.  The existence of such an extensive data set provides opportunities for addressing a multitude of research questions.  For example, we previously used these data to robustly establish the statistical validity and reliability of the instrument for a much broader student population than is typically possible for RBAs \cite{wilcox2016eclass}.  Additionally, ongoing research includes analysis of these data with respect to gender differences in E-CLASS scores \cite{wilcox2016gender}.  Using demographic data on the students along with meta-data on the course, we are able to examine gender differences while controlling for confounding variables (e.g., student major, course level).  Moreover, the size of the available data set allows for examination of the intersectional impact of different variables while still maintaining significant statistical power. 

Future work with this growing, national data set will utilize course meta-data collected by the automated system to investigate the impact of differences in pedagogy and laboratory structure on students' epistemologies about experimental physics.  Aggregation of data from multiple courses over multiple semesters at CU and a few other institutions can also be used to examine changes in students' epistemologies over time using longitudinal and pseudo-longitudinal data.  Ultimately, this data set may help to provide concrete suggestions for both instructors and researchers interested in the nature of students' epistemologies about experimental physics and how they change over time.  

\subsubsection{Faculty use of the E-CLASS results}

Overall, roughly 90 physics instructors have received one or more E-CLASS reports over the past five semesters.  To better understand how these faculty perceive and use these reports, we distributed a short email survey to a subset of these instructors (N=55) and received 17 responses.  We also performed follow-up interviews with 9 of these instructors to gain additional detail regarding their responses to the survey.  A significant theme from these interviews was related to the perceived value of having significant and appropriate national data with which to compare individual course results.  Instructors noted that this comparison data helped them to determine what questions and trends to pay attention to when interpreting their students' performance.  

The surveys and interviews also showed that the E-CLASS reports have been used by instructors in multiple productive ways.  For example, 9 of the 17 survey respondents reported using data from their report to inform specific changes to their lab courses, such as revised lab manuals, introduction of reflective assignments, and changes in the grading structure to explicitly target practices such as making predictions and basing conclusions on data.  In several cases, these changes were made in the context of larger scale laboratory course transformation efforts involving multiple courses and instructors.  Additionally, 13 of 17 survey respondents reported discussing their E-CLASS report with colleagues or department chairs, and roughly a third of these respondents ($N=4$) included their results in presentations at local or national professional meetings.  Two faculty members also included their E-CLASS report as part of their tenure or post-tenure evaluation.  These findings suggest that the E-CLASS and its associated report have impacted faculty and their laboratory teaching practice in multiple ways across multiple institutions.  Moreover, with the increased sustainability afforded by the automation of both the administration and report generation process, the E-CLASS system can continue to serve as a resource for new and returning instructors.

\subsection{\label{sec:other}Other centralized assessment systems}

In addition to the E-CLASS, there are several other systems that exemplify some or all aspects of a centralized model of RBAs (Fig.\ \ref{fig:models}b), including the PhysPort Data Explorer \cite{PhysPortDE} and the Learning Assistant Alliance's Learning Assistant Supported Student Outcomes (LASSO) system \cite{lasso2015website}.  While a comprehensive comparison of the E-CLASS, PhysPort, and LASSO systems is beyond the scope of this work, this section briefly describes key feature of these two system.

The PhysPort Data Explorer is designed to be a tool for scoring, analyzing, and interpreting results of RBAs \cite{PhysPortDE}.  The system is currently in the open beta-testing phase, allowing individual instructors to upload their students' responses to a variety of RBAs.  The Data Explorer is not designed to be a fully centralized assessment system as it does not assist with administration or data collection.  Rather, this system represents a halfway point between the default model and a fully centralized model.  In this halfway model, data collection is still primarily in the hands of the instructors, but the system will provide a mechanism for aggregating centralized comparison data and offer tools for standardized analysis.  Unlike the E-CLASS system, where the analysis functions generally require that the data were collected directly by the system, the Data Explorer will also provide the opportunity for instructors to analyze data from supported RBAs that they have already collected from their courses.  

Recently, the Learning Assistant (LA) Alliance's has also created a new centralized, online assessment system to support their LASSO study \cite{vandusen2015lasso}.  Like the PhysPort Data Explorer, the LASSO system offers a range of RBAs; however, similar to the E-CLASS system, the LASSO system both hosts the RBAs and manages data collection, in addition to facilitating scoring and analysis.  To use the LASSO system, an instructor does not need to use learning assistants in their course, but they must join the LA Alliance (which is free).  Participating instructors register their course with the LASSO system and upload a course roster with email addresses for each student.  The system then emails a unique link to the RBA to each student on the roster.  After administering the post-test, instructors also receive a summary document reporting their students' pre- and post-test distributions and averages.  

One consistent difference between the E-CLASS system and that of the Data Explorer and LASSO is that both the Data Explorer and LASSO are designed as a hub for multiple RBAs, whereas the E-CLASS system offers only the E-CLASS.  RBA developers interested in utilizing a centralized assessment model for new or existing RBAs will be able to decide if they wish to create a new system for their assessment or incorporate it into an existing, multiple-RBA system.  Creating a new system affords a greater degree of control (e.g., access to raw data) and customizability (e.g., customized instructor reports), but also requires significant investment of time and funding to create and maintain the system.  Utilizing an existing multi-RBA system eliminates the burden of developing and maintaining a new system, but potentially affords less opportunity to customize the administration and reporting for the particular assessment.

\section{\label{sec:discussion}Discussion}

The E-CLASS system, along with the PhysPort Data Explorer and the LA Alliance's LASSO system, demonstrates that a centralized model for the dissemination of RBAs represents a viable and potentially sustainable alternative to the passive dissemination model that has been used historically in PER.  Centralized assessment systems have two primary advantages over this default model.  The first advantage is increased instructor support for administration, analysis and interpretation of results.  Particularly since RBAs are typically given at the beginning and end of the course, it can be very difficult for busy instructors to find the time and resources to administer an extra assessment.  By managing the survey generation and data collection process, centralized systems greatly reduce the burden placed on instructors.  Moreover, particularly for instructors with minimal experience with RBAs, the analysis and representation of RBA data to facilitate interpretation of the results can also present a significant barrier.  By providing standardized analysis and reporting, centralized assessment systems provide a mechanism for instructors to take advantage of the collective expertise of the PER community with respect to analyzing, representing, and presenting the results of RBAs, rather than having to develop this expertise for themselves.  

The second major advantage of a centralized assessment model is the potential for aggregating large data sets of responses from a more diverse population of students, institutions, and courses.  These large-scale data sets can then be used both to provide comparison data and for research purposes.  Comparison data is a key tool for instructors trying to productively interpret their students' performance.  Without such comparison data, it can be difficult for instructors to make clear and actionable conclusions about their students' performance.  National data sets from RBAs also represent a valuable resource for physics education researchers interested in broader research questions related to student learning outcomes, particularly when students' responses are accompanied by detailed information about course context and pedagogy like that collected on the E-CLASS Course Information Survey.  

Despite the clear advantages offered by centralized assessment systems, these systems also have a number of drawbacks and limitations.  For example, the development of these systems is time consuming and expensive, and they require some degree of maintenance.  Each of the systems described in this paper (E-CLASS, LASSO, and Data Explorer) had external funding from the National Science Foundation (NSF) and/or another organization.  For the E-CLASS system, the majority of this funding was used to cover the initial development costs of the system, including hardware, software, and personnel.  However, even for automated systems, some minimal funding to support a part-time administrator will likely be necessary to respond to user issues and maintain the system indefinitely.  (Note that this funding was in addition to the initial funding required to support development and validation of the assessment.)  Another potential limitation of the centralized assessment systems discussed here is that they have been tested only with multiple-choice assessments.  Indeed, the administrative viability of these systems is arguably dependent on the system being able to score students responses electronically.  The ability to consistently and reliably score open-ended questions electronically has yet to be demonstrated within PER.  Thus, at the moment, centralized assessment systems may be feasible only for RBAs with forced-choice question formats.  

The E-CLASS and other similar systems have now demonstrated that a model for dissemination of research-based assessments that centralizes administration, data collection, and analysis represents a viable alternative to the default, historical model.  We argue that this model has significant advantages and recommend that members of the PER community give legitimate consideration to this alternative model when developing and disseminating research-based assessments in the future.  Research-based assessments are an important component of furthering efforts to improve physics education, and centralized assessment systems have the potential to facilitate the use of these assessments for both instructors and physics education researchers.

\begin{acknowledgments}
This work was funded by the NSF-IUSE Grant DUE-1432204 and NSF Grant PHY-1125844.  The statements made in this article do not necessarily represent the views of the NSF.  Particular thanks to Takako Hirokawa and to the members of PER@C for all their help and feedback.  
\end{acknowledgments}

\bibliography{master-refs-ECLASS-11-15}

\end{document}